\documentclass[12pt]{iopart}
\usepackage{iopams}  
\expandafter\let\csname equation*\endcsname\relax
\expandafter\let\csname endequation*\endcsname\relax
\usepackage{amsmath,amsthm, amssymb,amsfonts}
\usepackage{rotating}

\usepackage[english]{babel}
\usepackage{graphicx}

\usepackage{t1enc}
\usepackage[T1]{fontenc}
\usepackage[latin1]{inputenc}
\usepackage{url}
\usepackage{amsfonts}
\usepackage{setspace}

\begin{document}

\title[Validation of EEG forward modeling approaches in the presence of anisotropy]{Validation of EEG forward modeling approaches in the presence of anisotropy in the source space}

\author{F.~Drechsler$^{1}$, J.~Vorwerk$^{2}$, J.~Haueisen$^{3}$, L.~Grasedyck$^{4}$, and C.~H.~Wolters$^{2}$}

\address{\begin{itemize}
    \item [$^1$] Max Planck Institute for Mathematics in the Sciences, Inselstr. 22, D-04103 Leipzig, Germany    
    \item [$^2$] Institute for Biomagnetism and Biosignalanalysis, University of M{\"u}nster, 48149 M{\"u}nster, Germany
    \item [$^3$] Institute of Biomedical Engineering and Informatics, TU Ilmenau, POB 100565, D-98684 Ilmenau, Germany
    \item [$^4$] Institute for Geometry and Applied Mathematics, RWTH Aachen, Templergraben 55, D-52056 Aachen, Germany
\end{itemize} 
}
\ead{carsten.wolters@uni-muenster.de}

\begin{abstract}
The quality of the inverse 
approach in electroencephalography (EEG) source analysis is --- among other things --- 
depending on the accuracy of the forward modeling approach, 
i.e., the simulation of the electric potential for a known dipole
source in the brain.\\ 
Here, we use multilayer sphere modeling scenarios to investigate the 
performance of three different finite element method (FEM)
based EEG forward approaches -- subtraction, Venant and partial integration -- 
in the presence of tissue conductivity anisotropy in the source space. 
In our studies, the effect of anisotropy on the potential is 
related to model errors when ignoring anisotropy and to numerical errors, 
convergence behavior and computational speed of the different FEM approaches. 
Three different source space anisotropy models that best represent adult, child and 
premature baby volume conduction scenarios, are used.\\ 
Major findings of the study include (1) source space conductivity anisotropy has a
significant effect on electric potential computation: The effect increases with increasing 
anisotropy ratio; (2) with numerical errors far
below anisotropy effects, all three FEM approaches are able to model source space 
anisotropy accordingly, with the Venant approach offering the best 
compromise between accuracy and computational speed;  
(3) FE meshes have to be fine enough in the subdomain between the source 
and the sensors that capture its main activity.
We conclude that, especially for the analysis of cortical development,
but also for more general applications using EEG source analysis techniques, 
source space conductivity anisotropy should be modeled and the  
FEM Venant approach is an appropriate method.

\end{abstract}



\vspace{2pc}

\maketitle

\section{Introduction}
In EEG and magnetoencephalography (MEG) source analysis, 
the inverse solution, i.e., the reconstruction of a current distribution 
in the brain from non-invasive measurements of the electric potential 
at the head surface, depends --- among other things --- on the accuracy of the 
forward problem~\cite{CHW:Bre2012,CHW:Hae93,CHW:Luc2012}. 
For an appropriate solution of the forward problem it is important to keep 
both {\em model} and {\em numerical errors} as small as possible. 

{\em Model errors} result, for example, from unrealistic or simplifying assumptions 
on head volume conductor geometry and tissue conductivity structure. In this paper,
we will focus on the model error caused by ignoring tissue conductivity
anisotropy in the source space (the brain grey matter compartment). 
After the volume conductor model has been determined, 
{\em numerical errors} of the chosen numerical method result from an insufficient 
discretization and approximation of the potential 
due to, for example, insufficient mesh resolution, bad (deformed) elements 
or an inappropriate choice of the basis functions with regard to the 
regularity of the solution function.

Different numerical methods have been proposed with the goal
to solve the EEG and MEG forward problem in the presence of brain tissue conductivity 
anisotropy, under which are the finite difference method (FDM)~\cite{CHW:Hal2005,CHW:Nei2005}, the finite volume 
method (FVM)~\cite{CHW:Coo2008} and the finite element method (FEM). This paper 
focuses on the FEM which allows high accuracy for the numerical solution of elliptic partial differential
equations since it is specifically tailored to the corresponding
variational formulation~\cite{CHW:Hac92,CHW:Bra2007,CHW:Wol2007e,CHW:Dre2009} and since it 
allows high flexibility and accuracy in modeling the forward problem 
in geometrically complicated inhomogeneous and anisotropic head volume 
conductors~\cite{CHW:Ber91,CHW:Awa97,CHW:Bro97,CHW:Buc97,CHW:Mar98,CHW:Oll99,CHW:Wei2000,CHW:Sch2002,CHW:Wen2008,CHW:Val2010,CHW:Gue2010,CHW:Pur2011,CHW:Pur2012b}.

Recent studies investigated the influence of brain anisotropy in realistic head 
models~\cite{CHW:Hau2002,CHW:Wol2005a,CHW:Wol2006,CHW:Hal2008b,CHW:Gue2010}. 
It was found that, besides the non-negligible effects of the so-called 
{\em remote anisotropy} (anisotropy in a certain distance to the source),
conductivity changes close or within the source space 
have an especially significant effect on the forward problem~\cite{CHW:Wol2005a,CHW:Hau2000}. 
However, to the best of our knowledge, a validation of FEM-based forward approaches 
in the presence of anisotropy in the source space has not yet been presented, 
and the relationship between model errors due to anisotropies in the source space
and numerical errors remains unclear. The standard in source analysis is still the use
of a homogenized isotropically conducting source space.  

When using the effective medium approach, developed and validated 
in~\cite{CHW:Bas94,CHW:Tuc2001,CHW:Oh2006,CHW:Wan2008}, that relates 
conductivity tensors and water diffusion tensors in a linear way, brain conductivity anisotropy 
can be determined non-invasively by means of diffusion tensor magnetic resonance 
imaging (DT-MRI or DTI) techniques. 

Using modern 7T MRI machines, evidence for grey matter 
anisotropy in adults was recently found in~\cite{CHW:Hei2010,CHW:Coh2012}. 
In~\cite[Fig. 4]{CHW:Hei2010}, a fractional 
anisotropy (FA) map with overlaid vector-orientation showed a clear normally-oriented 
(radial) anisotropy in the cortex. In~\cite[Figure 39.7]{CHW:Jon2010}, this result is confirmed. 
It has thus been measured that the largest diffusion 
tensor eigenvalue is the one in radial direction. In~\cite[Fig. 4]{CHW:Coh2012}, 
FA values in the range of up to 0.15 were measured for the area of the central sulcus, 
which would relate to an anisotropy ratio of about 1.3:1.
FA-values were not constant in the region of interest; higher FA was measured in Broadman area BA4 
(primary motor cortex) and lower in BA2 (sensory cortex). 
In \cite{CHW:Shi99}, an anisotropy ratio of up to 1.41:1 (in the head of the caudate nucleus) 
was measured for adults. This ratio would relate to an FA value of approximately 0.21.
This motivates our choice of a 1.41:1
(radial:tangential) conductivity anisotropy ratio for the grey matter compartment in our
{\em adult model} described in detail later in this study. To the best of our knowledge, 
it is not likely that an even more pronounced source space anisotropy is realistic for the adult case.

However, the human brain undergoes a dramatic maturation during the first years of life. 
Both the grey and white matters change rapidly during the first two years of life
as indicated by the changes in T1 contrast of the grey versus white matter around 
6 to 9 months after birth~\cite{CHW:Alm2007}.
In the period 26 to 29 weeks of gestation, a rapid dendritic differentiation in the 
cortical plate, the appearance of a six-layered pattern and a maximal development of
the subplate zone has been reported~\cite{CHW:Mrz88}. 
As shown in~\cite{CHW:Nei2000,CHW:Nei2002}, the grey matter anisotropy ratio is 
increasing with decreasing age. Since the investigation of the cortical development using 
non-invasive EEG and MEG techniques is of increasing 
interest~\cite{CHW:Esw2002,CHW:Pre2004,CHW:Pih2004,CHW:Oka2006,CHW:Roc2008,CHW:Pih2009,CHW:Lew2011}, a correct modeling of
especially these higher levels of source space anisotropy might significantly contribute
to an understanding of the observed maturation effects. In~\cite{CHW:Nei2000}
and in~\cite[Fig.~2]{CHW:Nei2002}, anisotropy ratios have been measured in the 
cortex of premature babies of 26 weeks of gestational age, which could otherwise only been 
found in white matter tissue. Therefore, for our {\em premature baby} model described in detail
in the methods section of this paper, we assume a source space anisotropy 
ratio of 5:1 as a representation of an upper limit of grey matter anisotropy 
in humans and for our {\em child model} a ratio of 
2.7:1~\footnote{unpublished maximal cortical anisotropy ratio in a DTI study 
of a child}. 

The three chosen cortical anisotropy ratios are upper limits of grey matter anisotropy
and surely only rough approximations of the realistic situations, especially since they are known
to vary strongly depending on the exact location in the grey matter 
compartment~\cite{CHW:Shi99,CHW:Coh2012}. The chosen ratios have to be seen in the light 
of the main interest of this study: We want to investigate  
relationships of grey matter anisotropy modeling errors and 
corresponding numerical errors. The goal is to gain deeper insight 
in the effects of cortical conductivity anisotropy on the EEG forward problem
and to present the requirements with regard to a successful
FEM modeling. Only if we can find FEM parametrizations
which lead to numerical errors that are below the model 
errors, we can conclude that we model grey matter anisotropy accordingly. 
Since numerical errors can only be studied in simplified volume conductors, 
we restrict this study to multilayer-sphere models with constant 
radial and tangential conductivity values for the different 
tissue compartments, where quasi-analytical solutions exist~\cite{CHW:Mun93}. 

In the next section, we will shortly introduce the EEG forward problem and
three different FEM approaches for the modeling of a current dipole in 
inhomogeneous and anisotropic volume conductors, 
the {\em subtraction approach} described in detail in~\cite{CHW:Dre2009,CHW:Wol2007e}, 
the {\em Venant approach} introduced in~\cite{CHW:Buc97} and described in detail 
in~\cite{CHW:Wol2007a} and the {\em partial integration (PI)} approach~\cite{CHW:Wei2000,CHW:Val2010}.
In our previous work~\cite{CHW:Lew2009b}, we compared all three approaches
in tetrahedral volume conductors with isotropic source space, while,
for the first time, we validate in the presence of cortical conductivity 
anisotropy here. We then describe our 
measures for model errors (or cortical anisotropy {\em effect errors}) 
and numerical errors and detail the setup of the volume conductor models.
We optimize cortical conductivity values so that anisotropy effects
are purely due to anisotropy and not to a suboptimal choice of the corresponding
homogenized isotropic conductivity. Starting from a coarser mesh, 
we then generate globally and locally refined tetrahedral models
to examine convergence behavior of the different FEM approaches.
In the first two studies, the refined mesh models are constraint to fulfill 
the so-called {\em homogeneity condition} of the full subtraction 
approach, namely the condition that conductivity is not allowed to 
change in a small subdomain around the source (in our setup
the tetrahedron of the coarsest mesh that contains the source). 
The result section is divided in three studies: The first 
and the second study, where globally refined and locally refined
meshes are used, resp., have the goal to present the effect of cortical
conductivity anisotropy in the three different volume conductor models,
relate those effects to the numerical errors and determine the FEM 
approach that performs best with regard to numerical error, 
computational speed and constraints on mesh resolution and practical applicability. 
In the third study, the model error is further reduced by means
of relaxing the homogeneity condition and the best-performing FEM 
approach of studies 1 and 2 is used for computation. In the last section, 
we discuss results and conclude.
   
\section{Methods}
\subsection{Forward problem formulation}
Assuming the quasi-static approximation of the Maxwell equations, the electric potential $u$ in the head domain $\Omega$ with conductivity distribution $\sigma$, evoked by a primary current $\mathbf{j}^p$, is given by a Poisson equation with homogeneous Neumann boundary conditions on the head surface $\Gamma = \partial \Omega$. This can be expressed by  
\begin{equation} \label{pde_part1}
  \nabla \cdot (\sigma \nabla u ) = \nabla \cdot \mathbf{j}^p = J^p \mbox{ in } \Omega,  \qquad
  \langle \sigma \nabla u , \mathbf{n} \rangle = 0 \mbox{ on } \Gamma, 
\end{equation}
with $\mathbf{n}$ being the unit surface normal. To achieve 
uniqueness of the solution a reference electrode with given potential has to be fixed, 
i.e., $u(\mathbf{x}_{ref}) = 0$ \cite{CHW:Sar87,CHW:Hae93,CHW:Wol2007e}. 
A primary current at position $\mathbf{y} \in \mathbb{R}^3$ and with moment $\mathbf{M} \in \mathbb{R}^3$ 
is represented by a mathematical dipole
\begin{equation} \label{jpdelta}
J^p(\mathbf{x}) = \nabla \cdot \mathbf{j}^p(\mathbf{x}) = \nabla \cdot (\mathbf{M} \cdot \delta(\mathbf{x} - \mathbf{y} )),
\end{equation}
where $\delta$ is the Dirac delta distribution \cite{CHW:Sar87,CHW:Mun88b,CHW:Hae93,CHW:Wol2007e}.

\subsection{Finite element method}

We discretize \eqref{pde_part1} using a variational formulation 
and a Galerkin FE approach. We choose linear basis functions 
$\varphi_i$ at the vertex positions $\xi_i$, $i \in \{ 1 , \ldots , N \}$, 
of a discretization of $\Omega$. 
The finite element approach leads to the following linear equation system
\begin{equation} \label{equsys}
K u = b^\mathbf{y}, \quad K \in \mathbb{R}^{N \times N }, \quad u,b^\mathbf{y} \in \mathbb{R}^N
\end{equation}
\cite{CHW:Wol2007e,CHW:Dre2009} with the stiffness matrix 
\[
K_{ij} := \int_\Omega \langle \sigma (\mathbf{x}) \nabla \varphi_j(\mathbf{x}) , \nabla \varphi_i(\mathbf{x}) \rangle \: d\mathbf{x}, \quad i,j \in \{ 1, \ldots , N \},
\]
and the right-hand side
\begin{equation} \label{pde_rhs}
b^\mathbf{y} := - \int_\Omega J^p(\mathbf{x}) \varphi_i(\mathbf{x}) \: d\mathbf{x} = - \int_\Omega \nabla \cdot \mathbf{j}^p(\mathbf{x}) \varphi_i(\mathbf{x}) \: d\mathbf{x}.
\end{equation}

\subsection{Dipole models}
\label{subsec:dipmod}
As described in the introduction, we evaluate three different FE approaches for the treatment 
of the singularity on the right hand side (\ref{pde_rhs}) with regard to their performance in the 
presence of source space anisotropy. In the following, we will give a short overview of their 
derivations.
\subsubsection{Subtraction approach}
Under the assumption that there exists a non-empty open neighborhood $\Omega_{\infty}$ of the source 
position $\mathbf{y}$ with constant (possibly anisotropic) conductivity 
$\sigma^{\infty,\mathbf{y}} := \sigma(\mathbf{y}) \in \mathbb{R}^{3 \times 3}$, we 
split up the potential $u$ and the conductivity $\sigma$ into the parts
\begin{eqnarray} 
u &= u^{corr,\mathbf{y}} + u^{\infty,\mathbf{y}},\\
\sigma &= \sigma^{corr,\mathbf{y}} + \sigma^{\infty,\mathbf{y}}.
\end{eqnarray}
Here, 
$u^{\infty,\mathbf{y}}$, the so-called {\em singularity potential}, is the analytical solution for a dipole
in an unbounded homogeneous conductor with constant (possibly anisotropic) conductivity 
$\sigma^{\infty,\mathbf{y}}$, which can be computed as
    \begin{equation}\label{u_infty}
      u^{\infty,\mathbf{y}}(\mathbf{x})  
      := 
      \frac{1}{4\pi\sqrt{{\rm det}\,\sigma(\mathbf{y})}}
      \frac{\langle \mathbf{M}(\mathbf{y}),\sigma(\mathbf{y})^{-1}(\mathbf{x}-\mathbf{y})\rangle}{
        \langle \sigma(\mathbf{y})^{-1}(\mathbf{x}-\mathbf{y}),\mathbf{x}-\mathbf{y}\rangle^{3/2}}.
    \end{equation}
$u^{corr,\mathbf{y}}$ is the so-called {\em correction potential} that solves the equation
\begin{align} \label{pde2}
  \nabla \cdot (\sigma \nabla u^{corr,\mathbf{y}} ) = f(\mathbf{x}) \mbox{ in } \Omega, \\
  \langle \sigma \nabla u^{corr,\mathbf{y}} , \mathbf{n} \rangle = g(\mathbf{x}) \mbox{ on } \Gamma
\end{align}
with
\begin{align}
  f(\mathbf{x}) &:= \nabla ( \sigma (\mathbf{y}) - \sigma (\mathbf{x})) \nabla u^{\infty,\mathbf{y}} (\mathbf{x}) \quad \text{for} \; \mathbf{x} \in \Omega\\
  g(\mathbf{x}) &:= - \langle \sigma (\mathbf{x}) \nabla u^{\infty,\mathbf{y}} (\mathbf{x}) , \mathbf{n}(\mathbf{x}) \rangle \quad \text{for} \; \mathbf{x} \in \Gamma
\end{align}
and is computed numerically. Applying the FE method, we obtain the right hand side
\begin{align}
  b_i^\mathbf{y} &:= \int_\Omega \langle ( \sigma (\mathbf{y}) - \sigma (\mathbf{x}) ) \nabla u^{\infty , \mathbf{y} } (\mathbf{x}) , \nabla \varphi_i (\mathbf{x}) \rangle d\mathbf{x}  \nonumber \\ 
  & - \int_\Gamma \varphi_i(\mathbf{x}) \langle \mathbf{n}(\mathbf{x}) , \sigma (\mathbf{y}) \nabla  u^{\infty , \mathbf{y} } (\mathbf{x}) \rangle d\Gamma, \quad i \in \{ 1, \ldots, N\} \label{rhssub}
\end{align}
and, after solving \eqref{equsys} with right hand side \eqref{rhssub} for the correction potential, we obtain the approximated solution for the total potential 
\[
u(\mathbf{x}) = \sum_{i = 1}^N (u_i + u^{\infty,\mathbf{y}}( \xi_i) )\varphi_i(\mathbf{x}).
\]
We refer to~\cite{CHW:Dre2009,CHW:Wol2007e} for more detailed mathematics of the subtraction approach,
where a proof for existence and uniqueness and statements about convergence properties
can be found.

\subsubsection{Partial integration approach}
Applying partial integration to \eqref{pde_rhs} gives us 
\begin{align*}
  - \int_\Omega (\nabla \cdot \mathbf{j}^p) \varphi_i \: d\mathbf{x} = \int_\Omega \langle \mathbf{j}^p, \nabla \varphi_i \rangle \: d\mathbf{x} - \int_\Gamma \langle \mathbf{j}^p , \mathbf{n} \rangle \: d\Gamma.
\end{align*}
We exploit the fact that the current density vanishes on the head surface and achieve
\begin{equation}
  - \int_\Omega (\nabla \cdot \mathbf{j}^p) \varphi_i \: d\mathbf{x} = \int_\Omega \langle \mathbf{j}^p , \nabla \varphi_i \rangle \overset{(\ref{jpdelta})}{= } \langle \mathbf{M} , \nabla \varphi_i (\mathbf{y} ) \rangle.
\end{equation}
This gives us the right hand side for the partial integration approach
\begin{equation}
  b_i^\mathbf{y} := 
  \left\{ 
  \begin{array}{ll} 
    \langle \mathbf{M} , \nabla \varphi_i (\mathbf{y} ) \rangle & \mbox{if } i \in \textit{NodesOfEle}(\mathbf{y}), \\
    0 & \textit{otherwise}.
  \end{array} 
  \right.
\end{equation}
where $\textit{NodesOfEle}(\mathbf{y})$ gives the nodes of the finite element that contains the dipole.

\subsubsection{Venant approach}
Exploiting Saint Venant's principle, we approximate a point dipole by a distribution of electrical monopoles \cite{CHW:Buc97,CHW:Wol2007a}. We place monopoles at the finite element node closest to the source position and those sharing an edge with this node. We adjust the charges of the monopoles so that the dipole moment of this configuration matches the dipole moment of the source as good as possible, while the sum of charges and higher moments remain zero. To avoid numerical instabilities as a consequence of huge source loads, an additional regularization is needed \cite{CHW:Buc97,CHW:Wol2007a}. If we denote the set of neighboring FE nodes for a dipole at position $\mathbf{y}$ by $\textit{Neighbours}(\mathbf{y})$ and the strength of the monopole at FE node $\xi_i \in \textit{Neighbours}(\mathbf{y})$ by $q_i$, we obtain the right-hand side
\begin{equation}
   b_i^\mathbf{y} := 
  \left\{ 
  \begin{array}{ll} 
    -q_i & \mbox{if } \xi_i \in \textit{Neighbours}(\mathbf{y}), \\
    0 & \textit{otherwise}.
  \end{array} 
  \right.
\end{equation} 

\subsection{Computational efficiency}
The two direct approaches Venant and PI, which approximate the dipolar source locally 
by a distribution of electrical monopoles, have a sparse right hand side and allow 
the calculation of a single solution in a few milliseconds when applying the transfer 
matrix approach \cite{CHW:Wol2004}. In contrast, the subtraction approach 
has a much higher computational demand since both the right hand side computation \eqref{rhssub} 
and also its multiplication to the transfer matrix are computationally
demanding since \eqref{rhssub} is a fully populated vector (see~\cite{CHW:Dre2009}
for more details).

\subsection{Quasi-analytical solution, difference criterion and electrode configuration}
In \cite{CHW:Mun93}, a quasi-analytical solution was derived for a mathematical dipole in an anisotropic multilayer sphere model that will be used here to compute modeling errors (isotropic versus anisotropic)
and numerical errors (numerical versus quasi-analytical).
We use
the relative error (RE)
\[
RE(\mathbf{u}^1,\mathbf{u}^2):= \frac{\| \mathbf{u}^1 - \mathbf{u}^2 \|_2}{\| \mathbf{u}^2 \|_2}
\]
as a difference measure throughout our study. 
$\mathbf{u}^1,\mathbf{u}^2\in\mathbb{R}^m$ with $m$ the number of surface electrodes might, for example, denote the numerical and the quasi-analytical solutions or the forward solution in anisotropic and isotropic brain compartment models, resp..

To achieve error measures which are independent of the choice of the sensor configuration, we distribute the electrodes in a most-regular way over the outer sphere surface. In this way we generated a configuration with $m=748$ electrodes on the surface of the outer sphere of our four layer sphere model.

\subsection{Volume conductor models}
\label{subsec:volcondmod}
Three four-layer sphere volume conductor models are used
in our study, an {\em adult model}, a {\em child model} and a {\em premature baby model}.
 
\begin{table}[t]
  \begin{center}
    \caption{Four layer sphere models used for the isotropic reference scenarios} \label{table_layers}
    \begin{tabular}{ccc}
      \hline
      Compartment            & Outer radius  & Isotropic conductivity (S/m)    \\ \hline
      Skin                   &   92mm        &    0.43            \\ 
      Skull                  &   88mm        &    0.01            \\ 
      CSF                    &   80mm        &    1.79           \\ \hline
      Brain (adult)          &   76mm        &    0.33            \\ 
      Brain (child)          &   76mm        &    0.51            \\ 
      Brain (premature baby) &   76mm        &    0.59            \\ \hline
    \end{tabular}
    
  \end{center}
\end{table}
\begin{table}[t]
  \begin{center}
    \caption{Conductivities in radial and tangential directions used for the anisotropic brain scenarios} \label{tableaniso}
    \begin{tabular}{ccc}
      \hline
      Model    & radial (S/m) & tangential (S/m)\\ \hline 
      Adult     &  0.41   & 0.29      \\ 
      Child      &  0.783  & 0.29      \\ 
      Premature baby  &  1.2    & 0.24      \\ \hline
    \end{tabular}
  \end{center}
\end{table}
Table \ref{table_layers} presents the radii and the conductivity parameters for the isotropic reference scenarios of the four spherical layers that approximate the compartments 
skin, skull, cerebrospinal fluid (CSF) and brain. While the radii of all compartments and
the isotropic conductivities of the compartments skin~\cite{CHW:Ram2006}, 
skull~\cite{CHW:Dan2011} and CSF~\cite{CHW:Bau97} were fixed throughout our studies, 
the conductivity of the brain compartment differs between the three models 
in the isotropic (Table \ref{table_layers}) and anisotropic (Table \ref{tableaniso}) modeling situations.

For the adult model in Table \ref{table_layers}, we used the standard isotropic 
brain conductivity of $0.33 S/m$ as generally used in source analysis~\cite{CHW:Hal2005,CHW:Nei2005,CHW:Ram2006}. 
Furthermore, we fixed the ratios between radial and tangential conductivities in 
Table \ref{tableaniso} for the anisotropic brain models of 1.41:1 (adult), 
2.7:1 (child) and 5:1 (premature baby) following the motivation given in the 
introduction. 

The values for radial and tangential conductivity for the adult model in Table \ref{tableaniso}
were computed by means of an optimization procedure. We computed forward solutions
using the quasi-analytical forward method for 16,338 dipole locations and 
three dipole moments in the carthesian directions, overall thus $49,014$
forward solutions, first for the isotropic adult reference scenario and, 
in a second step, for a series of corresponding 
anisotropic scenarios with fixed ratio of 1.41:1 but differing magnitudes  
for radial and tangential conductivities. For each of those anisotropic scenarios,
we summed up the RE's between isotropic and anisotropic electrode potentials
over all $49,014$ sources. The radial and tangential conductivity parameters 
with the minimal RE were then chosen as the corresponding anisotropic 
brain adult model as indicated in Table \ref{tableaniso}. 
We can therefore state that the presented effects in the results section
are {\em anisotropy effects} and not due to suboptimally chosen 
conductivity values. 

We now motivate the chosen optimized conductivity parameters for 
the child and premature baby models. The grey matter cytoarchitecture 
in premature babies is dominated by the apical dendrites of 
pyramidal cells that have few branches and are arrayed 
radially~\cite{CHW:Nei2000,CHW:Nei2002}. The decrease of 
relative anisotropy in the grey matter compartment
(in our model a decreasing anisotropy ratio from 5:1 to 2.7:1 and down to 1.41:1) 
was accompanied by a decrease in the average apparent diffusion 
coefficient values (in our model an assumed decreasing radial conductivity 
in Table \ref{tableaniso} and an assumed decreasing homogenized isotropic 
conductivity in Table \ref{table_layers}) when development 
progresses~\cite{CHW:Nei2000,CHW:Nei2002}. 
The authors furthermore report that during maturation
pyramidal cells elaborate highly-branched basal dendrites which tend to be arranged 
parallel to the cortical surface and, in addition, thalamocortical 
afferent fibers are added that are also oriented parallel to the 
cortical surface~\cite{CHW:Nei2000,CHW:Nei2002}.
We therefore assumed a slightly increasing tangential conductivity
between premature baby and child model (Table \ref{tableaniso}). 
Under these constraints, we performed again an optimization
as explained for the adult model to ensure that the chosen homogenized
isotropic conductivities for child and premature baby model
in Table \ref{table_layers} have minimal RE's to
their corresponding anisotropic models shown in Table \ref{tableaniso}
so that the presented effects in the results section are again
{\em anisotropy effects} and not due to suboptimally chosen 
corresponding conductivity values.

\subsection{Tetrahedral mesh generation}
\label{chapter_mesh_generation}
\begin{figure*}
  \begin{center}
    \includegraphics[width=5cm]{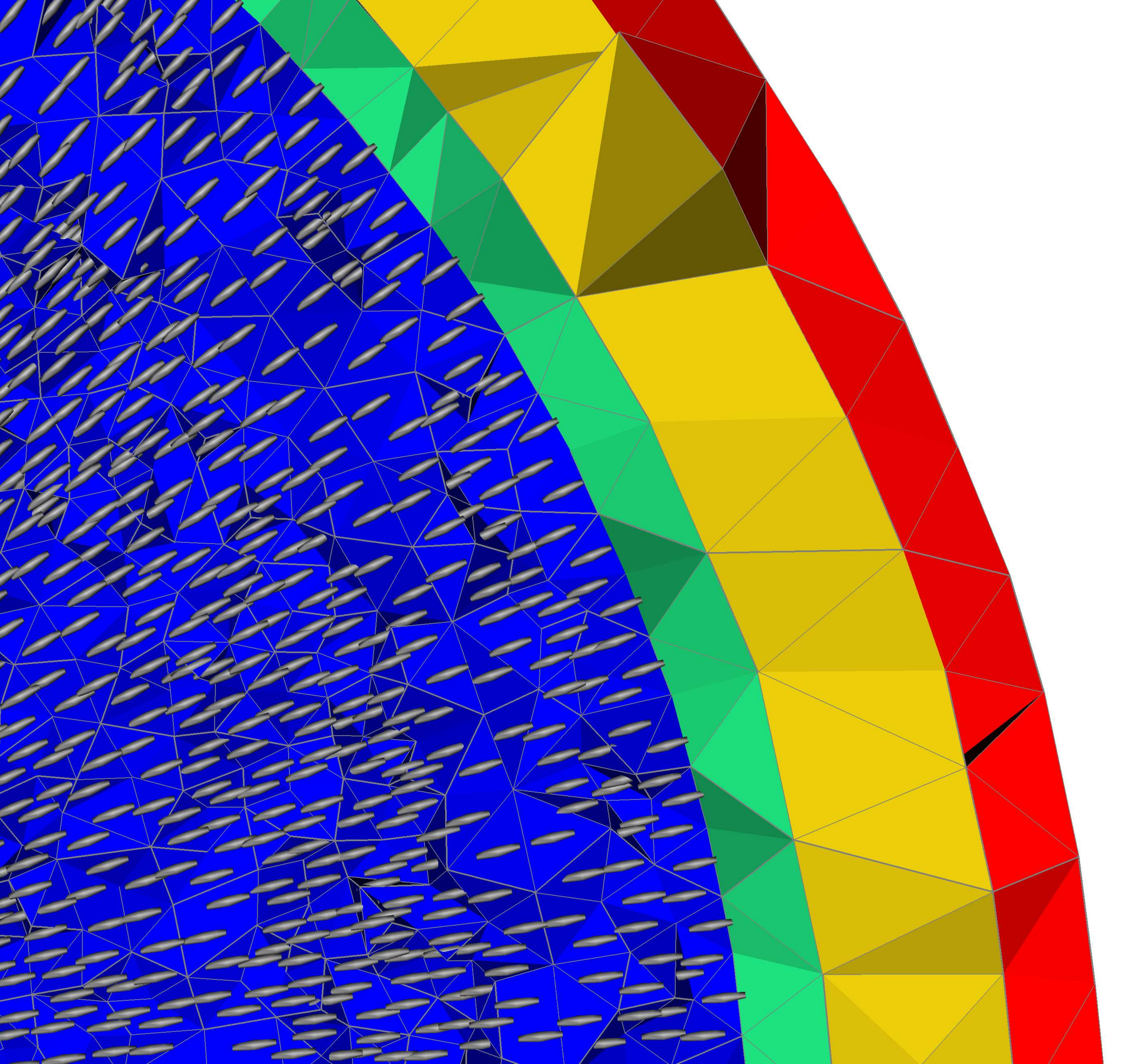}
    \includegraphics[width=5cm]{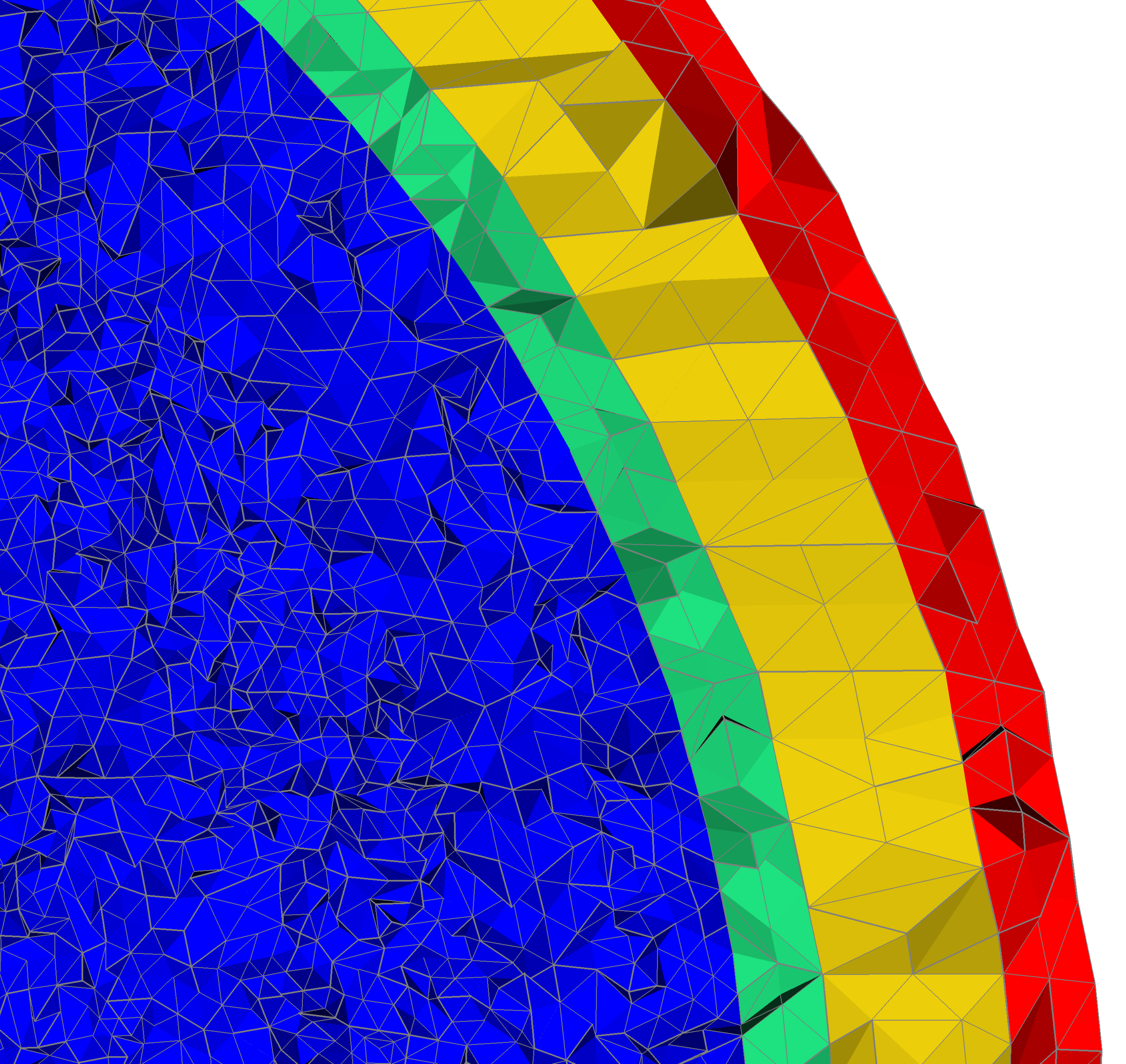}
    \includegraphics[width=5cm]{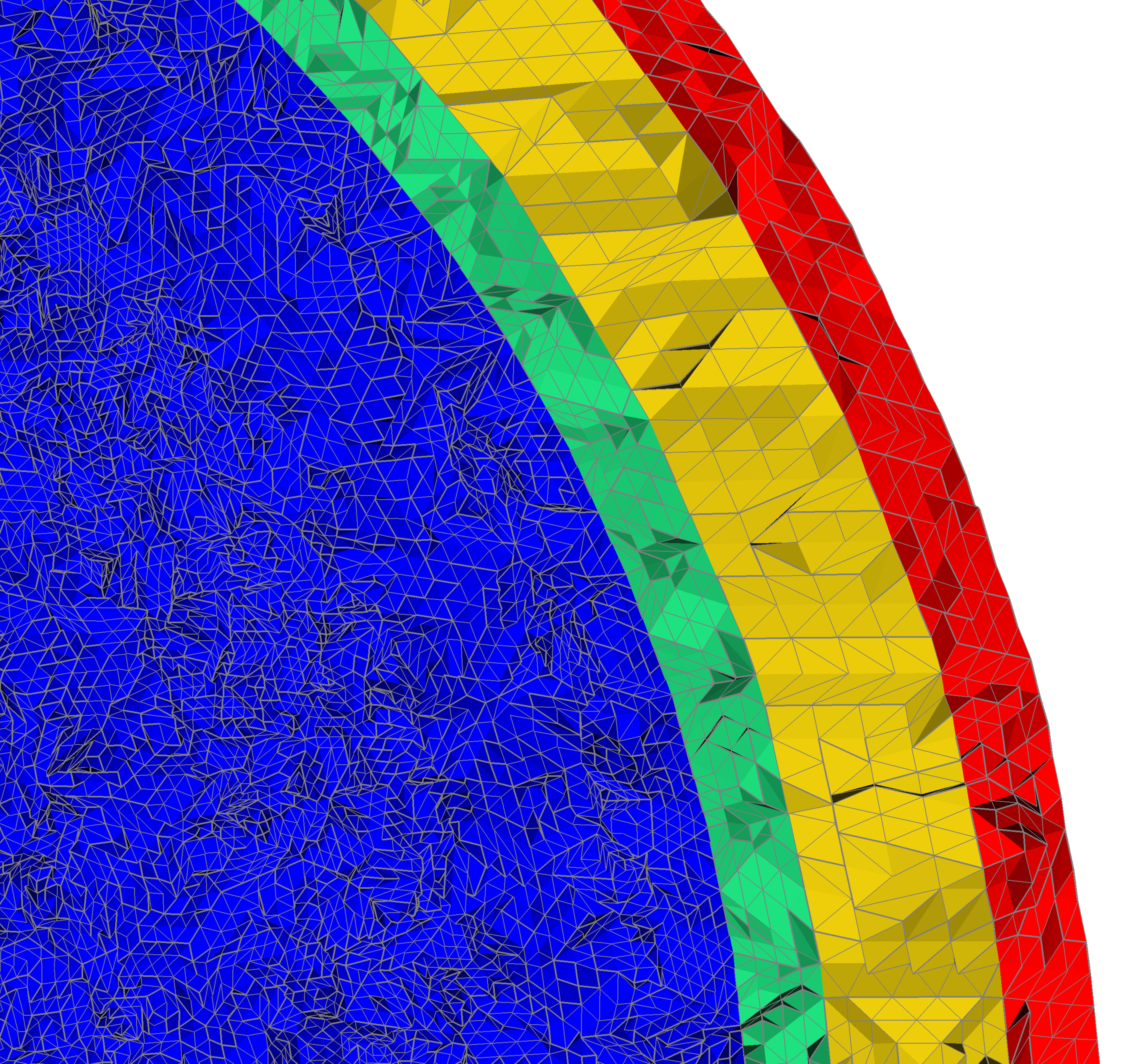}
  \end{center}
\caption{{\bf Visualization of the basic FE mesh {\em mesh142k} (left) and the globally-refined meshes {\em mesh1.1m} (middle) and {\em mesh9.5m} (right). See also Table~\ref{regulargrids}.}}
\label{fig_vis_meshes}
\end{figure*}
\begin{table}[t]
  \begin{center}
    \caption{Globally refined tetrahedral FE meshes: Generation procedure and number of nodes and elements.} \label{regulargrids}
    \begin{tabular}{cccc}
      \hline
      Mesh     & Generated with & Nodes         & Elements    \\ \hline 
      mesh142k & Tetgen         & 142,563       & 900,372     \\ \hline 
      mesh1.1m & Refinement     & 1,188,329     & 7,202,976   \\ 
      mesh9.5m & Refinement     & 9,590,961     & 57,623,808  \\ \hline
    \end{tabular}
  \end{center}
\end{table}
For the generation of our coarsest tetrahedral FE mesh, namely {\em mesh142k} 
(Table \ref{regulargrids} and Figure~\ref{fig_vis_meshes}), we used TetGen \cite{CHW:Tetgen}, which implements a 
constrained Delaunay tetrahedralization (CDT) approach \cite{CHW:Si2008,CHW:TetView}. 
The meshing procedure started with the preparation of a suitable
boundary discretization of the model. For each of the
four layers and for a given triangle edge length, nodes were
distributed in a most-regular way and connected through triangles.
This yielded a valid triangular surface mesh for each of the four layers.
Meshes of different layers were not intersecting each other. The CDT
approach was then used to construct a tetrahedralization conforming
to the surface meshes. It first built a Delaunay tetrahedralization of the
vertices of the surface meshes. It then used a local degeneracy removal
algorithm combining vertex perturbation and vertex insertion to
construct a new set of vertices which included the input set of vertices.
In the last step, a fast facet recovery algorithm was used to construct
the CDT~\cite{CHW:Si2005,CHW:Si2008}. This approach is combined with two 
further constraints to the size and shape of the tetrahedra. The first 
constraint can be used to restrict the volume of the generated 
tetrahedra in a certain compartment, the so-called {\em volume constraint}. 
The second constraint is important for the generation of quality 
tetrahedra. If $R$ denotes the radius of the unique circumsphere of a 
tetrahedron and $L$ its shortest edge length, the so-called 
radius-edge ratio of the tetrahedron can be defined as $Q = \frac{R}{L}$.
The radius-edge ratio can detect almost all badly-shaped tetrahedra
except one type of tetrahedra, so-called slivers. A sliver is a very flat
tetrahedron which has no small edges, but can have arbitrarily large
dihedral angles (close to $\pi$). For this reason, an additional mesh
smoothing and optimization step was used to remove the slivers and
improve the overall mesh quality.

\begin{table}[t]
  \begin{center}
    \caption{Locally refined tetrahedral FE meshes: Generation procedure, refinement area around the fixed 
             source location (in mm) and number of nodes and elements.} \label{locgrids}
    \begin{tabular}{ccccc}
      \hline
      Mesh        & Generated       & Ref. area       & Nodes         & Elements   \\
                  &   with          & around     &               &            \\ 
                  &                 & source    &               &            \\ \hline 
      mesh142k    & Tetgen          &    0mm                  & 142,563          & 900,372  \\  
      mesh142k\_10 & Refinement     &    10mm                 & 144,137           & 909,747       \\
      mesh142k\_20 & Refinement     &    20mm                 & 153,325           & 964,809       \\
      mesh142k\_30 & Refinement     &    30mm                 & 174,468           & 1,093,566       \\
      mesh142k\_40 & Refinement     &    40mm                 & 211,729           & 1,315,947       \\
      mesh142k\_50 & Refinement     &    50mm                 & 266,499           & 1,646,015       \\
      mesh142k\_60 & Refinement     &    60mm                 & 339,627           & 2,087,099       \\
      mesh142k\_70 & Refinement     &    70mm                 & 430,432           & 2,634,985       \\
      mesh142k\_100 & Refinement     &    100mm               & 774,189           & 4,710,009       \\
      mesh1.1m    & Refinement     & globally refined      & 1,188,329     & 7,202,976   \\ 
    \end{tabular}
  \end{center}
\end{table}
Based upon {\em mesh142k}, we created further globally and locally refined meshes. 
Refinement is carried out using regular refinement which splits 
every tetrahedron of the mesh in eight new tetrahedra~\cite{CHW:Bey98,CHW:Bey95}. 
The specialty of this approach is that the set of nodes of the coarser meshes 
are a subset of the nodes of each finer mesh, needed later for maintaining
the homogeneity condition of the subtraction approach. In this way, two regularly and globally
refined meshes were created (Table \ref{regulargrids} and Figure~\ref{fig_vis_meshes}).
Furthermore, we are interested in the effect of local in comparison 
to global mesh refinement. Therefore, we fixed the source
at \[(183.044;169.393;149.969)^T\] having a distance of 
73.93mm to the midpoint and of 2.07mm to the CSF. 
We then refined {\em mesh142k} inside a given radius around this position. First, a regular refinement, 
also called {\em red refinement}, of each tetrahedron inside the radius is applied. 
Secondly, in the {\em green refinement}, a closure is created so that the locally 
refined mesh has no hanging nodes~\cite{CHW:Bey98,CHW:Bey95} (Table~\ref{locgrids}).
We refine only \emph{mesh142k} and denote the locally refined meshes with 
\emph{mesh142k\_x} whereby x is the radius of the refinement area around the dipole. 
For example, if we refine all tetrahedra that lie in at most 10mm distance 
to the fixed position, we denote the corresponding model \emph{mesh142k\_10}.

For {\em mesh142k}, the conductivity tensors indicated in Tables \ref{table_layers}
and \ref{tableaniso} were then computed in the barycenters of the elements. 
First, with regard to the studies in the results-sections \ref{subsec:globrefmeshhom} 
and \ref{subsec:locrefmeshhom}, where we want to maintain the homogeneity condition 
for the subtraction approach, we assigned the conductivity of the original tetrahedron to each of the eight new 
tetrahedra in the refinement, that is, we accept that the refinement
will only improve the FE approximation, while the model error is not 
reduced. Secondly, for the study in results-section~\ref{subsec:globrefmeshadapt}, we calculate 
adapted conductivity tensors for the refined meshes, that is, we obtain 
meshes where also the conductivity tensors of the refined brain elements change from element 
to element.

\subsection{Dipole positions}
\label{subsec:dippos}

Dipoles were positioned in the brain compartment between the radii
68mm and 75mm, that is, in a distance of minimally 1mm and maximally 8mm 
to the CSF (the next bigger change in conductivity). We chose this intervall,
because it is well-known that numerical errors increase with decreasing distance
of the source to the next bigger conductivity jump (i.e., the 
CSF) and a mathematical reasoning for this phenomenon was given in~\cite{CHW:Wol2007e}.
We thus used on purpose a source space intervall where we have to be 
aware of bigger numerical forward modeling errors~\cite{CHW:Wol2007e,CHW:Dre2009,CHW:Lew2009b}. 
10 different dipole positions were chosen randomly in each of the
seven (68-69mm,...,74-75mm) 1mm intervals, that is, altogether 
70 different source positions.
Sources were placed in the barycenters of elements of 
{\em mesh142k}. Numerical errors were shown to be minimal 
for sources in element barycenters for both subtraction and PI approaches, 
while for the Venant approach, best source positions are the FE nodes~\cite{CHW:Vor2011}.
Since these errors are on an overall low scale~\cite{CHW:Vor2011}, our choice is slightly 
in favor of the subtraction and PI approaches.  
Since normally-oriented dipoles in the grey matter compartment are the 
anatomically and physiologically realistic sources in combination with the used 
radial conductivity anisotropy, we mainly only present results for 
radially-oriented sources. However, for academic interest,
we also present one result figure for tangentially-oriented sources.

\subsection{Computational platform and used software}
All simulations ran on an IBM server with 2 Xeon X5650 processors (2.67 GHz) using the SimBio software environment~\cite{CHW:SimBio}.

\section{Results}

For studies~\ref{subsec:globrefmeshhom} and \ref{subsec:locrefmeshhom}, 
the refined mesh models are constrained to fulfill the homogeneity condition of the full subtraction 
approach. The studies have the goal to present the effect of cortical
conductivity anisotropy in the three different volume conductor models,
relate those effects to the numerical errors and determine the FEM 
approach that overall performs best. 
In the last study~\ref{subsec:globrefmeshadapt}, the model error is further reduced by means
of relaxing the homogeneity condition, and the best-performing FEM 
approach of studies~\ref{subsec:globrefmeshhom} and \ref{subsec:locrefmeshhom} is used for computation. 

For the presentation of our results we use boxplots. 

Volume conductor model ({\em adult}, {\em child} or {\em premature baby}, see~\ref{subsec:volcondmod}) and 
the FE dipole modeling approach (Venant, PI or subtraction, see~\ref{subsec:dipmod}) are denoted in the title of each figure. The x-axis shows source eccentricity and the y-axis the used error measure.

\subsection{Globally refined meshes that fulfill the homogeneity condition}
\label{subsec:globrefmeshhom}
\begin{figure*}
  \begin{center}
    \includegraphics[width=5cm]{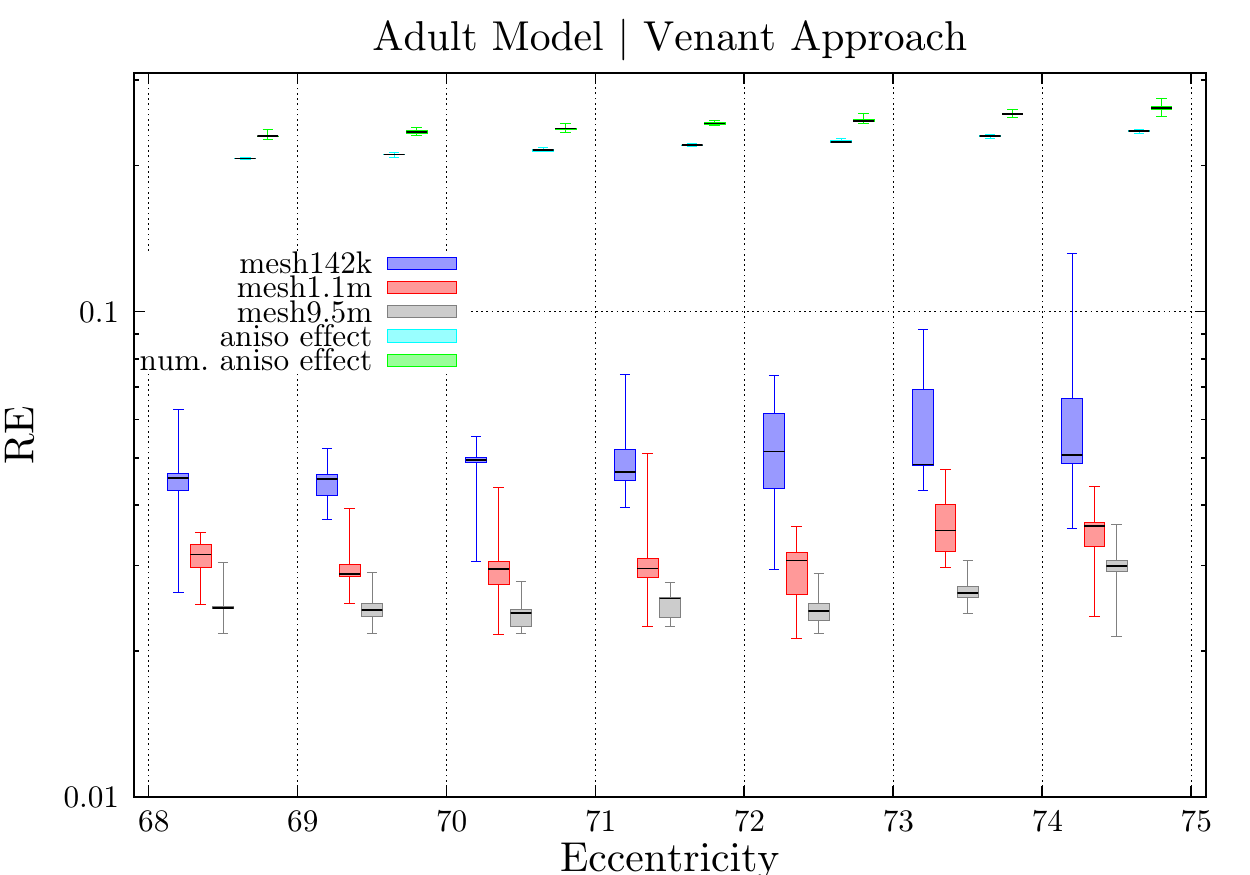}
    \includegraphics[width=5cm]{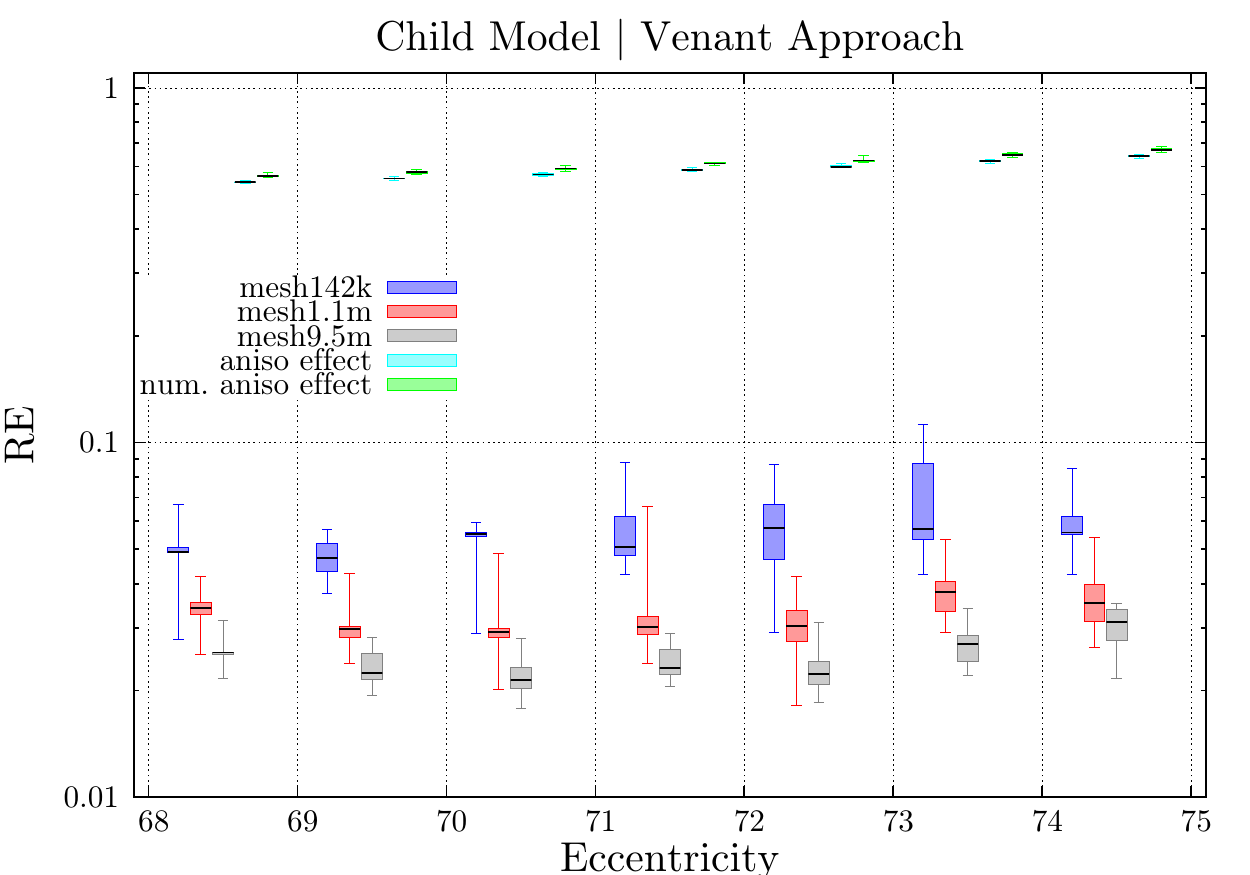}
    \includegraphics[width=5cm]{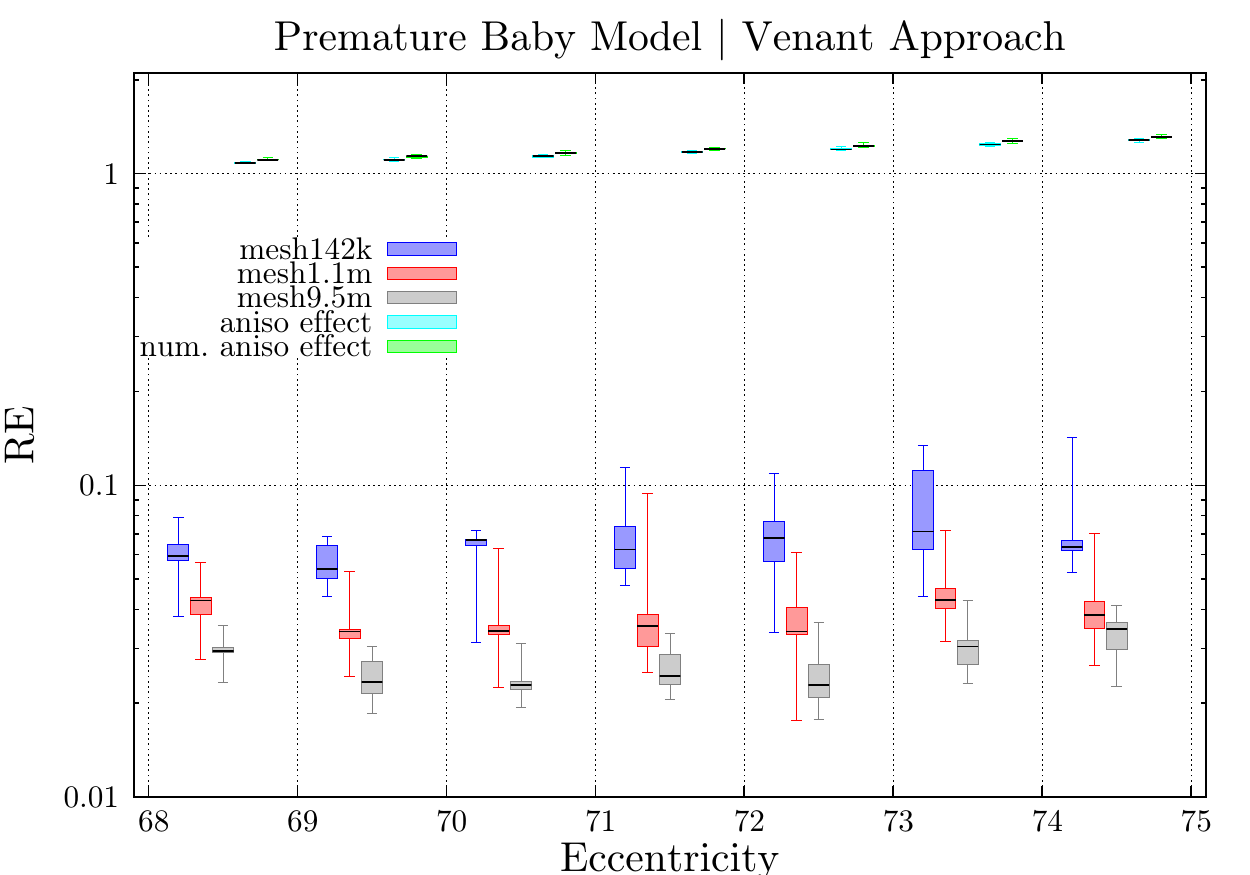}\\
  \end{center}
\caption{RE for normally-oriented sources and the Venant FE approach for {\em adult} (left column), 
{\em child} (middle column) and {\em premature baby} (right column) model: 
Boxplots are shown for the effect of cortical anisotropy (cyan), for the numerical anisotropy effect (green) 
(see description in~\ref{subsec:globrefmeshhom}) and for the numerical errors in the 
coarsest {\em mesh142k} (blue) and the once ({\em mesh1.1m}, red) and twice ({\em mesh9.5m}, grey) refined meshes.}
\label{fig_re_venant}
\end{figure*}
\begin{figure*}
  \begin{center}
     \includegraphics[width=5cm]{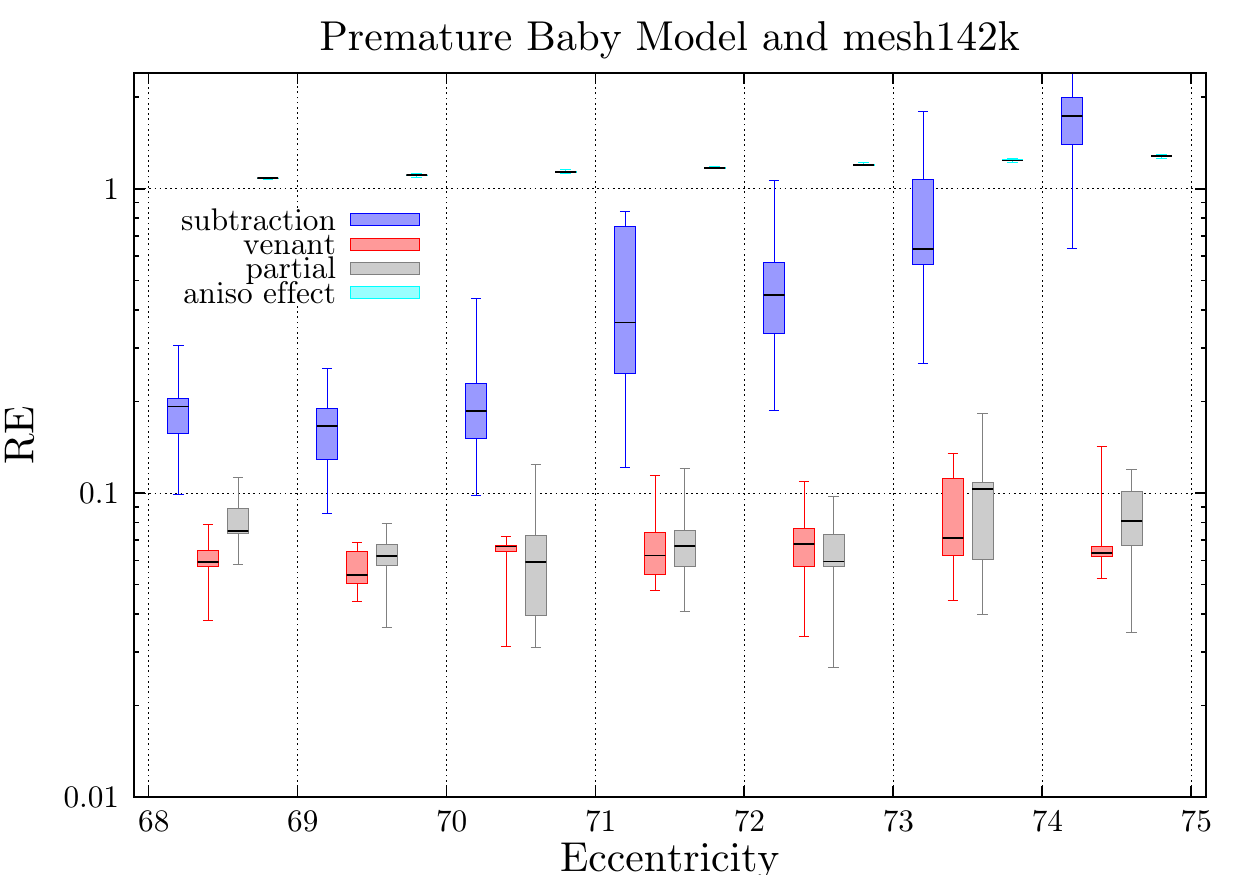}
     \includegraphics[width=5cm]{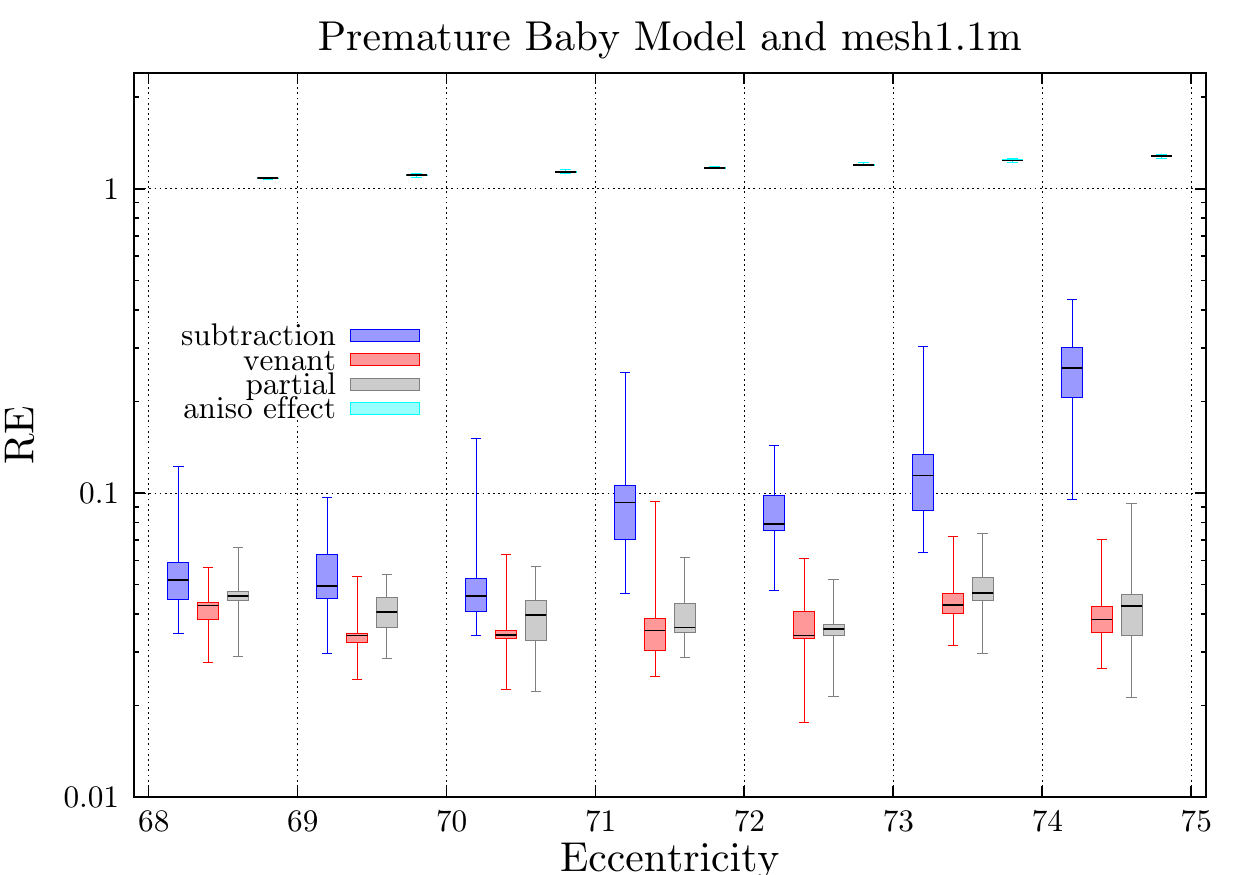}
     \includegraphics[width=5cm]{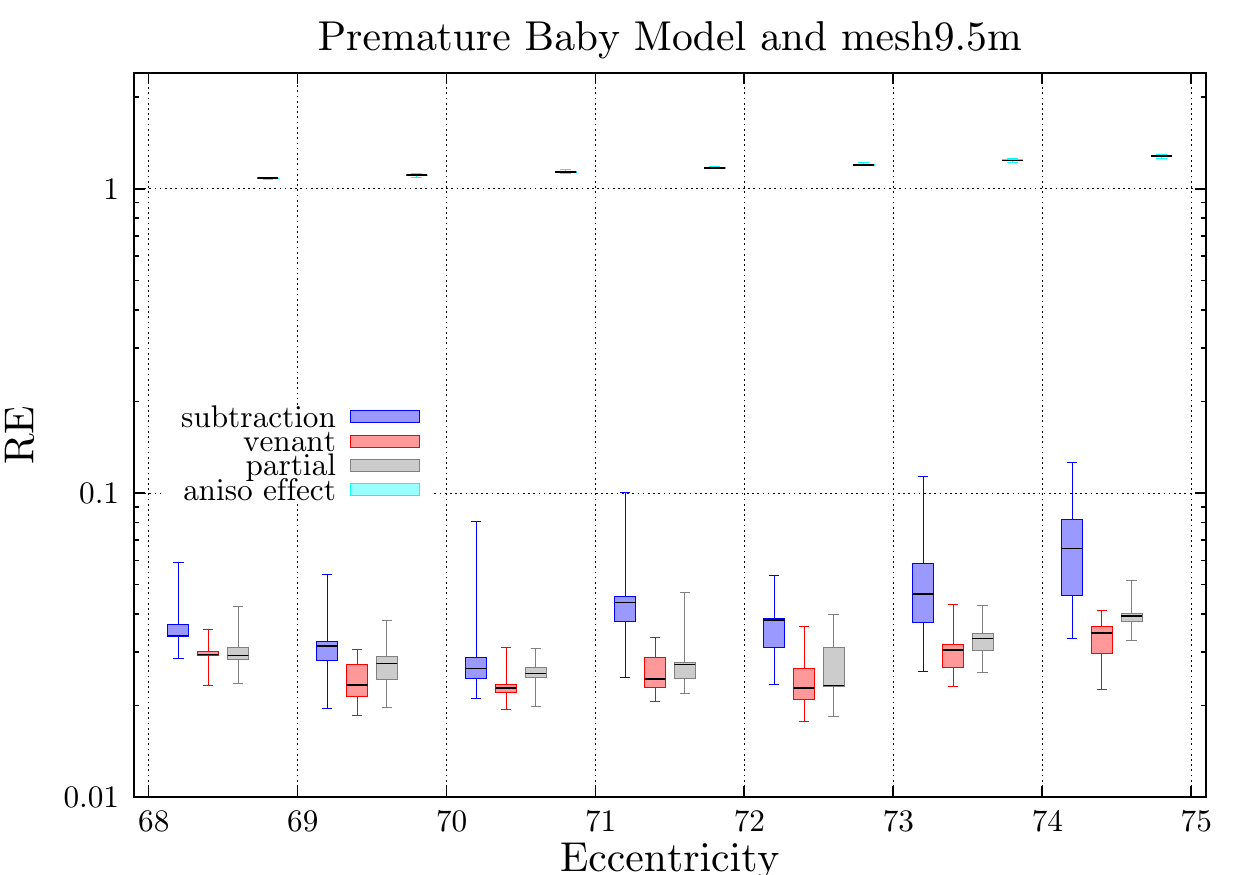}\\
     \includegraphics[width=5cm]{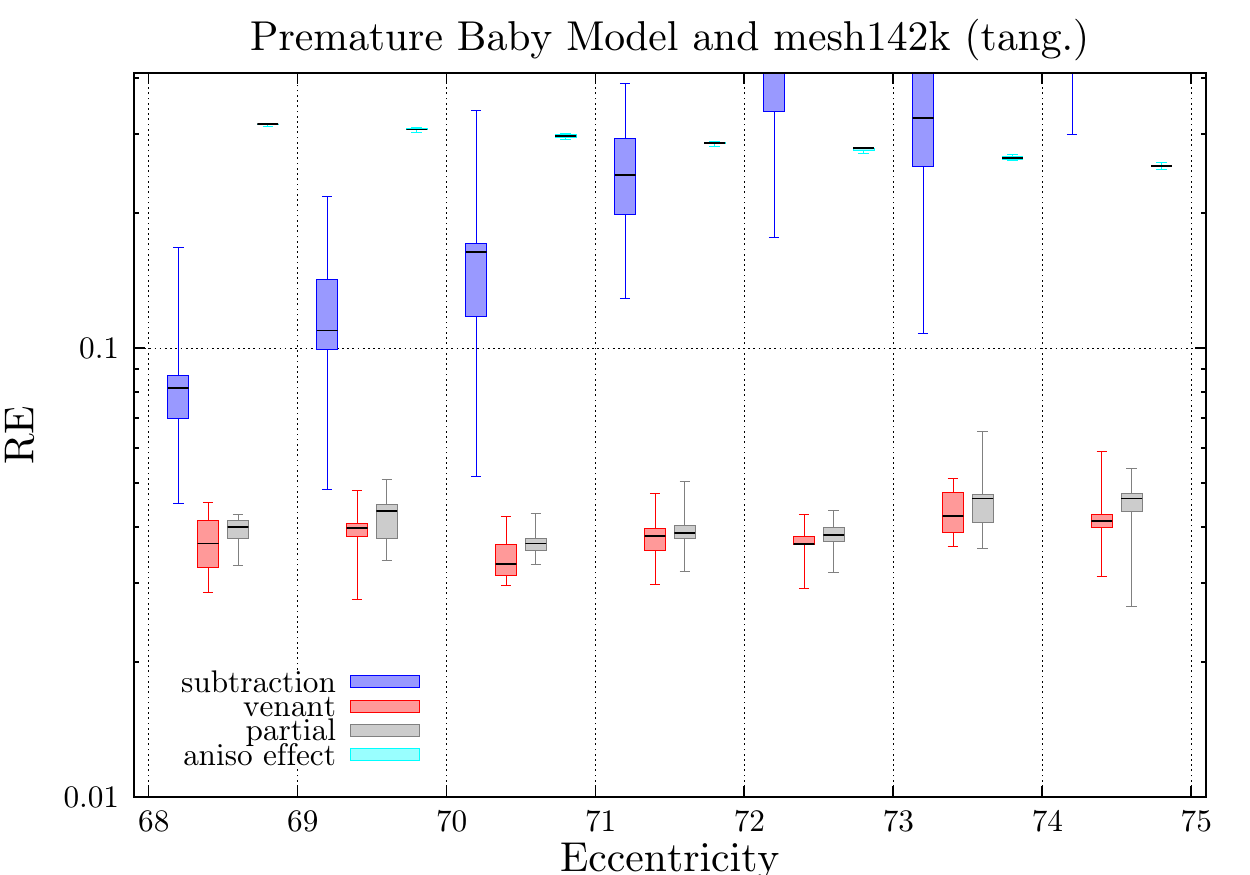}
     \includegraphics[width=5cm]{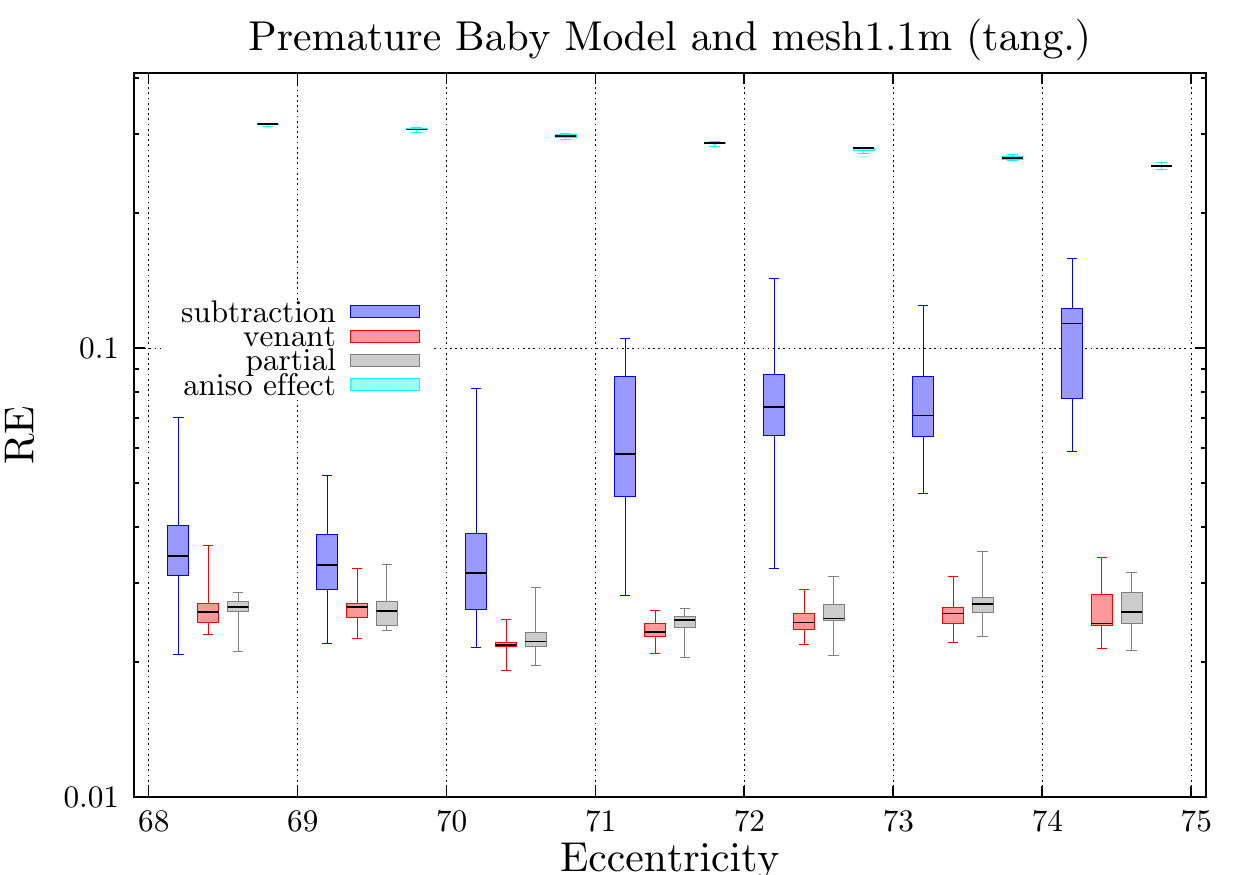}
     \includegraphics[width=5cm]{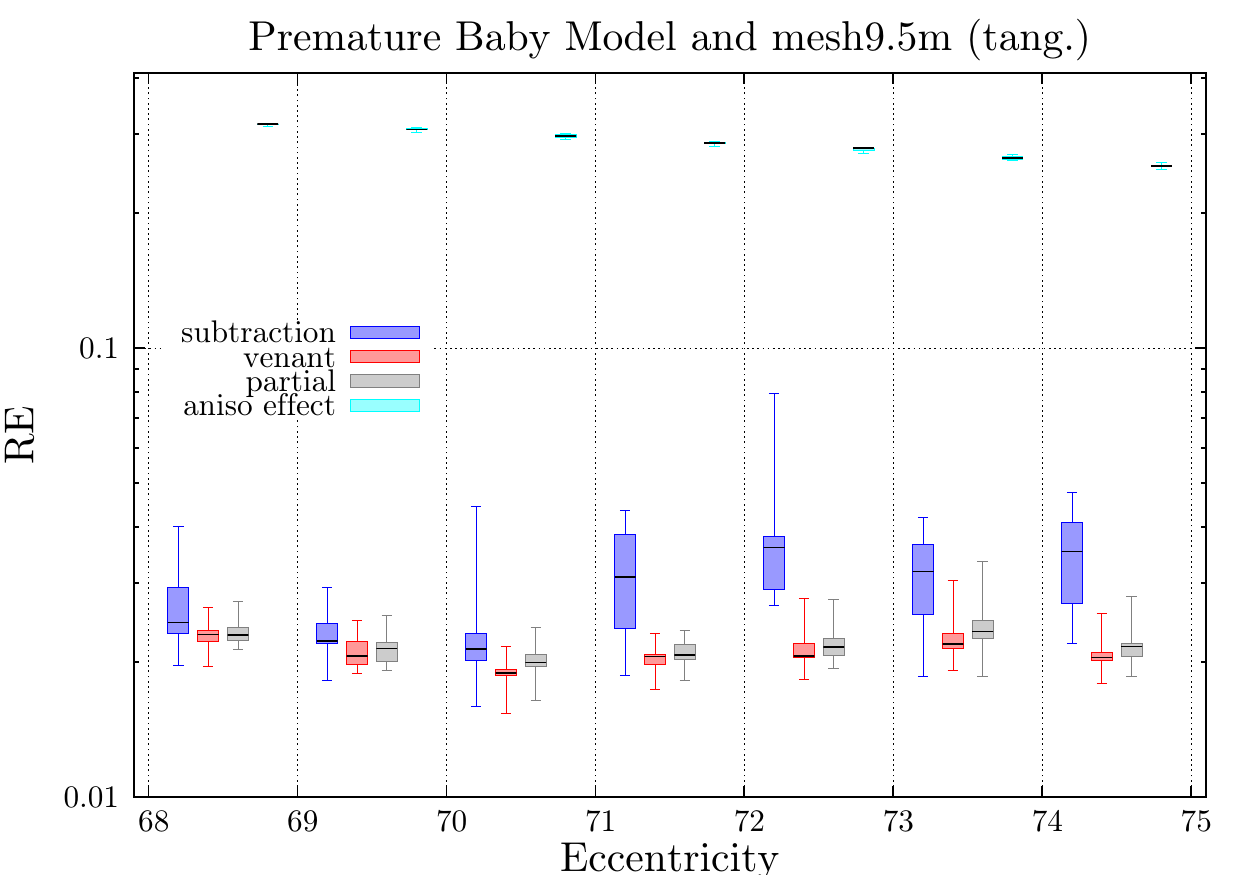}\\
  \end{center}
\caption{RE for normally-oriented (upper row) and tangentially-oriented (lower row) sources for the \emph{premature baby} model 
for the coarsest {\em mesh142k} (left column) and the once ({\em mesh1.1m}, middle column) and twice ({\em mesh9.5m}, right column) 
refined meshes. Boxplots are shown for the effect of cortical anisotropy (cyan) and for the numerical errors for subtraction (blue), 
Venant (red) and partial integration (grey) approach.}
\label{fig_re_compare}
\end{figure*}
%

%
Figures \ref{fig_re_venant} 
and \ref{fig_re_compare}
show the RE results for normally-oriented sources, 
for different volume conductor models and different dipole modeling approaches. 
The figures contain 
\begin{itemize}
\item
the effect of cortical anisotropy (\emph{aniso effect}, cyan), that is, the comparison between quasi-analytical forward results in the anisotropic and in the corresponding isotropic volume conductor model (see \ref{subsec:volcondmod}, Tables~\ref{table_layers} and \ref{tableaniso}), 
\item
the numerical anisotropy effect (\emph{num. aniso effect}, green), defined here as the results of 
the Venant approach in {\em mesh9.5m} for isotropic conductivities
versus the quasi-analytical results in the corresponding anisotropic model and 
\item
the numerical errors of the three different FE approaches in the coarsest {\em mesh142k} and 
in the once ({\em mesh1.1m}) and twice ({\em mesh9.5m}) refined meshes, that is, the numerical 
forward modeling results versus the quasi-analytical results, both for anisotropic conductivity 
in the cortex.
\end{itemize}

Figure \ref{fig_re_venant} presents those results for the Venant approach and for the 
{\em adult} model (left), for the {\em child} model (middle) and for the 
{\em premature baby} model (right). 
The \emph{aniso effect} (cyan) is increasing proportionally
to the degree of cortical anisotropy in the model, that is, it increases from
the {\em adult} (in intervall 68-69mm, an RE median of 0.2067)
to the {\em premature baby} model (an RE median of 1.0821).
As expected, the \emph{num. aniso effect} (green) closely follows the \emph{aniso effect}, only on
a slightly higher error scale, which shows that the error is mainly due to ignoring 
cortical anisotropy, while the additional numerical error is small.
Most importantly, for {\em mesh142k} (blue), {\em mesh1.1m} (red) and {\em mesh9.5m} (grey) 
we obtain a clear convergence behaviour, that is, the numerical errors
are still significant for the coarsest {\em mesh142k}, while for the finest mesh {\em mesh9.5m} 
they are in the area of 0.01$\le$ RE $\le$ 0.04 and thus far below the \emph{aniso effect}.

Figure~\ref{fig_re_compare} shows a direct comparison of the three FE approaches 
subtraction (blue), Venant (red) and partial integration (grey) in {\em mesh142k} 
(left column), {\em mesh1.1m} (middle column) and {\em mesh9.5m} (right column)
for both normally-oriented (upper row) as well as tangentially-oriented (lower row)
sources in the {\em premature baby} model. 
All three approaches show a clear convergence behavior with increasing mesh size. 
However, it is clearly visible that although the subtraction is by far the most computationally expensive FE approach
and although we chose source positions in favor of the subtraction (and the PI) approach,
it shows highest numerical errors for all three meshes. Both Venant and PI approach 
are at about the same error level with a sligh advantage of the Venant approach, even
if source positions were chosen in favor of the PI. 

\subsection{Locally refined meshes that fulfill the homogeneity condition}
\label{subsec:locrefmeshhom}
\begin{table*}
  \begin{center}
    \caption{Relative error (RE) for all three dipole models and locally refined meshes.}
    \begin{tabular}{c|ccc|ccc}
      \hline
      & \multicolumn{3}{c|}{normally-oriented sources} & \multicolumn{3}{c}{tangentially-oriented sources}\\ \hline
      Model & sub. & venant & partial & sub. & venant & partial                  \\ \hline 
      mesh142k       & 0.16899 & 0.07985 & 0.10579 & 0.09028 & 0.04279 & 0.04030\\ 
      mesh142k\_10   & 0.16844 & 0.07979 & 0.09093 & 0.09034 & 0.04286 & 0.04049\\ 
      mesh142k\_20   & 0.17036 & 0.07861 & 0.08995 & 0.09112 & 0.04446 & 0.04191 \\ 
      mesh142k\_30   & 0.18633 & 0.07101 & 0.07591 & 0.13560 & 0.03878 & 0.03075\\ 
      mesh142k\_40   & 0.05510 & 0.04554 & 0.04650 & 0.05632 & 0.02629 & 0.02507\\ 
      mesh142k\_50   & 0.06458 & 0.04153 & 0.04306 & 0.03438 & 0.02540 & 0.02354\\ 
      mesh142k\_60   & 0.06310 & 0.04132 & 0.04284 & 0.03377 & 0.02530 & 0.02318\\ 
      mesh142k\_70   & 0.06205 & 0.04154 & 0.04303 & 0.03578 & 0.02523 & 0.02305\\ 
      mesh142k\_100  & 0.06124 & 0.04198 & 0.04348 & 0.03354 & 0.02492 & 0.02278\\ 
      mesh1.1m       & 0.06048 & 0.04202 & 0.04344 & 0.03278 & 0.02492 & 0.02275\\ 
    \end{tabular}
    \label{table_local}
  \end{center}
\end{table*}
In the second study, we focus our interest on the numerical RE errors in 
the locally-refined meshes around the fixed source position as described 
in \ref{chapter_mesh_generation} (see, especially Table~\ref{locgrids}).
Table \ref{table_local} shows the results for all three dipole models
and for normally- and tangentially-oriented sources.
While the dipole has a distance of 18.07mm to the surface of 
the multisphere model, we observe a clear error decrease for a radius of 40mm
and nearly no more improvement for further refinement. Nearly the same
numerical accuracy could have thus been achieved with less than one fifth
of the FE nodes. Table \ref{table_local} confirms that the subtraction approach
can not compete against the direct approaches Venant and PI and, when taking
into account that the source position was chosen in the element barycenter,
that is, in favor of the PI approach, 
we can conclude 
again that the Venant approach seems to be the best choice out of the three
tested FE approaches.

\subsection{Globally refined meshes with adapted conductivities}
\label{subsec:globrefmeshadapt}
\begin{figure}
  \begin{center}
    \includegraphics[width=6cm]{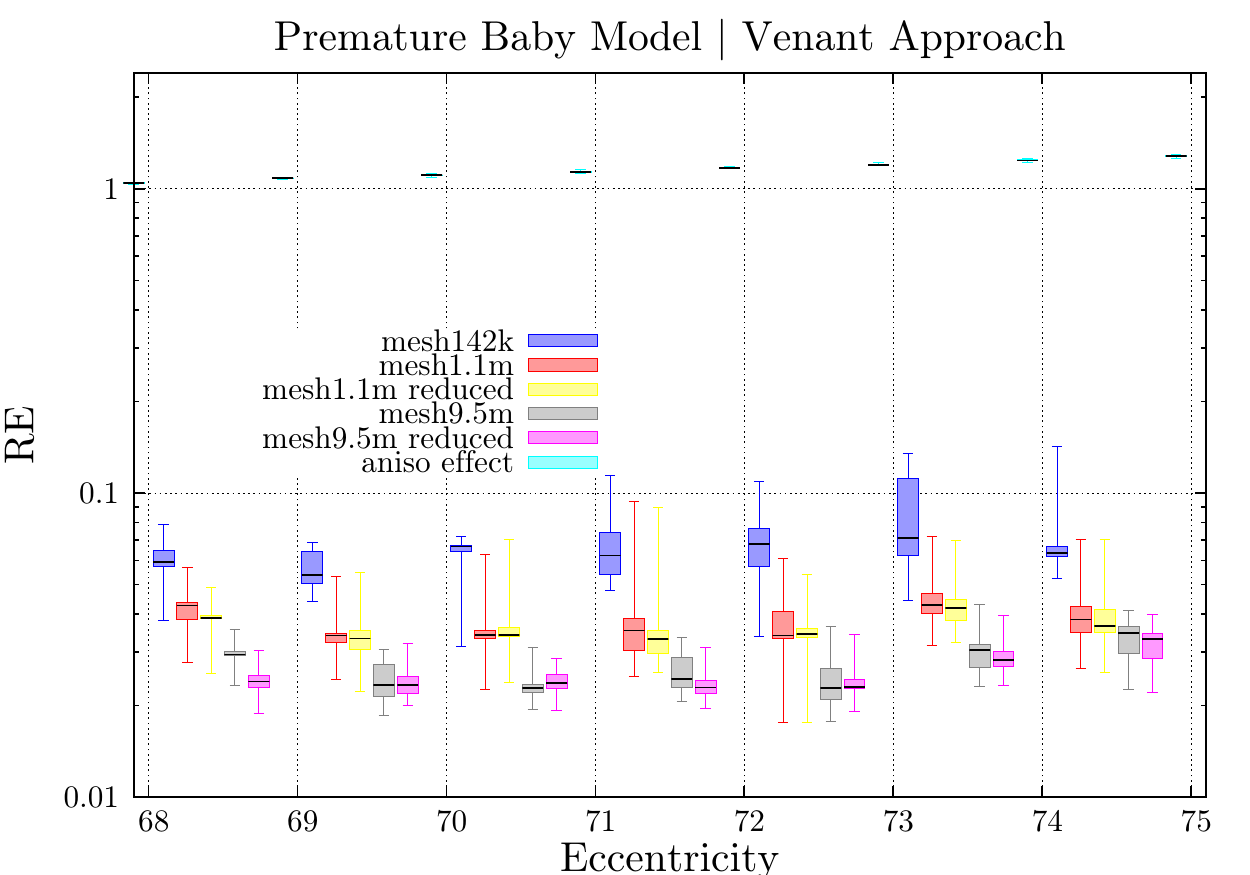} 
  \end{center}
\caption{RE for normally-oriented sources and the Venant FE approach for the \emph{premature baby} 
model and adapted conductivities: 
Boxplots are shown for the effect of cortical anisotropy (cyan) and for the numerical errors in the 
coarsest {\em mesh142k} (blue) and the once ({\em mesh1.1m}, red) and twice ({\em mesh9.5m}, grey) 
refined meshes without adapted conductivities from study \ref{subsec:globrefmeshhom} and for 
the once ({\em mesh1.1m reduced}, yellow) and twice ({\em mesh9.5m reduced}, pink)
refined meshes with adapted conductivities.}
\label{fig_reduced_model_error_RE}
\end{figure}
%
%
For the last study, we will only use the best performing FE approach from studies
\ref{subsec:globrefmeshhom} and \ref{subsec:locrefmeshhom}, namely the Venant
FE approach, and the model with highest cortical anisotropy, the premature baby model. 
The results of the last simulation study are shown in Figure \ref{fig_reduced_model_error_RE}. 
The presented boxplots for the \emph{aniso effect} (cyan) and for \emph{mesh142k} (blue) 
as well as for \emph{mesh1.1m} (red) and \emph{mesh9.5m} (grey) are the known 
numerical accuracy results for the globally-refined meshes from study \ref{subsec:globrefmeshhom}, 
that is, \emph{mesh1.1m} and \emph{mesh9.5m} fulfill the homogeneity condition. 
Additionally, we present the results in \emph{reduced mesh1.1m} (yellow) and \emph{reduced mesh9.5m} (pink),
where the homogeneity condition was relaxed, that is, where the conductivities 
in the refined models were adapted to the real model. A comparison
of the results \emph{mesh1.1m} versus \emph{reduced mesh1.1m} and \emph{mesh9.5m} versus
\emph{reduced mesh9.5m} shows only a slight overall RE numerical error decrease
for the two adapted models. The slight improvements in the modeling of the 
cortical conductivities thus does not show a significant improvement with regard to
the numerical errors.


\section{Discussion and conclusion}
In this paper, we studied the effect of cortical conductivity anisotropy versus modeling and numerical errors 
of three different FE approaches for the EEG forward problem, namely the subtraction, 
the partial integration (PI) and the Venant approach. To do so, we used multi-layer sphere volume conductors
and examined maximal cortical conductivity anisotropy for three different scenarios, reflecting the 
increasing anisotropy ratios from an {\em adult} over a {\em child} to a {\em premature baby} modeling 
situation. We assumed maximal anisotropy ratios of 1.41:1, 2.7:1 and 5:1, resp., as motivated
by the measurements of~\cite{CHW:Shi99,CHW:Nei2000,CHW:Nei2002}. 

We used a relative error (RE) based optimization procedure to find the best-approximating
conductivity parameters for the isotropy-anisotropy pairs that were compared. This procedure
allows us to state that the presented RE-effects are due to cortical anisotropy and not due to 
suboptimally chosen conductivity values. For other choices of the conductivities, the 
RE-errors would thus even be larger.

We showed that cortical conductivity anisotropy has a significant effect on the forward EEG problem
and should thus be modeled. The effect increased with increasing anisotropy ratio so that
the significance of modeling it increases with decreasing age of the studied subjects. 
Since the investigation of the cortical development is of high 
interest~\cite{CHW:Esw2002,CHW:Roc2008}, a correct modeling of
source space anisotropy might thus significantly contribute to an understanding of 
cortical maturation effects.
Realistic measurement of grey matter conductivity anisotropy
can be performed using a combination of newest MRI technology~\cite{CHW:Hei2010,CHW:Coh2012}
and the linear effective medium approach model for relating water diffusion and
conductivity~\cite{CHW:Tuc2001,CHW:Oh2006,CHW:Wan2008}, so that the application
of the proposed and evaluated methodology to realistic situations seems straightforward.

In our first two studies, we used globally- and locally-refined meshes that fulfilled 
the so-called {\em homogeneity condition}. This condition is important for the subtraction approach,
where the mathematical theory suggests that a 
sufficiently ({\em sufficient} relative to the mesh resolution) large neighborhood 
around the source position with constant conductivity is needed for an appropriate
treatment of the source singularity~\cite{CHW:Wol2007e,CHW:Dre2009}. 
In isotropically conducting brain compartments,
this can be easily and well fulfilled~\cite{CHW:Wol2007e,CHW:Dre2009}, but the situation 
is more complicated in volume conductors with anisotropic grey matter conductivity, 
as evaluated for the first time in this paper. We found that the smaller the neighborhood
(note that a smaller neighborhood offers the possibility of reducing the model error by means of adapting
the conductivities), the larger the numerical errors of the subtraction approach, 
resulting in larger overall errors. As a consequence,
for the subtraction approach, we found that it is better not to adapt conductivity tensors in refined
areas around the source. While the subtraction approach is very sensitive to this condition, it is
of less importance for both direct approaches PI and Venant, so that the performance
of those approaches could even slightly (even though, in our study with the chosen
parameters, not significantly) be improved (see study \ref{subsec:globrefmeshadapt}). 
This effect was only small, since the elements in the grey matter compartment were already 
relatively small for \emph{mesh142k}, so that already the coarsest mesh captured
the real conductivity structure in the volume conductor quite well.
More importantly, we showed FE convergence for all three FE approaches, so
that all of them are appropriate methods for the modeling of source space
anisotropy. Relative errors (RE) were shown to be far below 
anisotropy effect errors, especially for the highest FE resolutions.
We also investigated the relative difference measure (RDM) and the 
magnification factor (MAG)~\cite{CHW:Mei89} and found that 
the RE errors decomposed into both of them with, however, a more distinct MAG
component.
In our study with the chosen meshes, we found the Venant approach
to be the overall best-performing method with regard to accuracy, computational
complexity and practicability. We can thus state that cortical anisotropy should be modeled 
and, from the results of this study, the Venant FE approach should be 
used for it.

In study \ref{subsec:locrefmeshhom}, we showed that for accurate FE modeling, 
mesh-refinement is not only needed around the source, but 
between the source and those EEG sensors, that capture the main source activity. 
Since we did not observe an error reduction for larger refinement radii, we can also conclude that in areas far away from the source (for superficial
sources, e.g., deeper areas in the brain), coarser mesh resolutions seem sufficient to 
reduce computational complexity. However, since our FEM approach is completely linear 
in the number of nodes~\cite{CHW:Wol2004}, even high resolutions are practicable, as shown in this study,
where the finest resolution had nearly 10 Million FE nodes. 

Finally, we would like to mention that from our former experience when using realistic head 
models~\cite{CHW:Hau2000,CHW:Wol2005a,CHW:Wol2006} 
and from the results of this study, we expect that the proposed modeling is not only relevant for EEG,
but also for MEG, and, because of the higher anisotropy ratios, especially for modeling of 
newborn MEG~\cite{CHW:Pih2004,CHW:Oka2006,CHW:Pih2009,CHW:Lew2011} and 
fetal MEG~\cite{CHW:Esw2002,CHW:Pre2004}.

\section*{Acknowledgment}
This work was supported by the Deutsche Forschungsgemeinschaft (WO1425/3-1, GR3179/3-1, HA2899/14-1).

\section*{References}
\bibliographystyle{plain}
\bibliography{chw}

\end{document}